\documentclass[12pt]{iopart}
\usepackage{epsfig,harvard}
\newcommand{\isum}%
{\mathop{\hbox{$\displaystyle\sum\kern-13.2pt\int\kern1.5pt$}}}

\begin{document}
\letter{Calculation of the free-free transitions in the 
electron-hydrogen scattering S-wave model}
\author{Chris Plottke and Igor Bray\footnote{electronic address:
I.Bray@flinders.edu.au} 
}
\address{
Electronic Structure of Materials Centre,
The Flinders University of South Australia,
G.P.O. Box 2100, Adelaide 5001, Australia}
\date{\today}

\begin{abstract}
The S-wave model of electron-hydrogen scattering is evaluated using
the convergent close-coupling method with an emphasis on
scattering from excited states including an initial state from the
target continuum. Convergence is found for discrete excitations and the
elastic free-free transition. The latter is particularly interesting
given the corresponding potential matrix elements are divergent.
\end{abstract}
\maketitle

The convergent close-coupling (CCC) method has had many successes in the
field of electron-impact excitation and ionization of atoms and ions. 
In this method
the total wave function is expanded using $N$
square-integrable states and the close-coupling equations are solved
in the form of coupled
Lippmann-Schwinger equations for the $T$-matrix elements \cite{BS92}.
The  $N$ states are obtained from a truncated orthogonal Laguerre
basis, and thus in the limit as $N$ goes to infinity, the states
span the entire Hilbert space. 
The CCC method was tested by \citeasnoun{BS92l} on
the Temkin-Poet \cite{T62,P78} (S-wave) model of
electron-hydrogen scattering, where only states of
zero orbital angular momentum are retained. The total cross sections
for elastic, inelastic and ionization 
collisions converged, with increasing $N$, for all projectile energies
and agreed with the expected S-wave model solutions, where available. 

The success of the method for the S-wave model allowed application to
many real electron-atom scattering problems. However, application to
ionization processes revealed some fundamental difficulties
\cite{BF96,REBF97}, which have been subsequently best illustrated
by returning back to the S-wave 
model \cite{B97l}. Though the total ionization cross section (TICS)
was found to be convergent, the underlying singly differential cross
section (SDCS) was not necessarily so. The triplet SDCS showed rapid
convergence, but the singlet SDCS showed unphysical
$N$-dependent resonances. Furthermore, the SDCS were not found to be
symmetric about $E/2$, where $E$ is the total (excess) energy, even
though antisymmetry of the total wave function has been ensured
explicitly. It was suggested that for both total spin cases the
CCC($N$) amplitudes should converge (as
$N\to\infty$) to a step function,
being identically zero past $E/2$ \cite{B97l}. 
The step function model was attacked by
\citeasnoun{BC99} who claimed to have proved (see their Eq.(20)) that the 
CCC-calculated amplitudes should converge to the true amplitudes as
$N\to\infty$, and hence yield symmetric SDCS.
This claim was rebutted \cite{B99reply} and a number of
counterexamples given \cite{B99BR,B99ajp}. 

Unfortunately, a proof for the step function idea has not been given,
only suggestive numerical evidence provided. This has encouraged
others to study the problem more closely. \citeasnoun{BRIM99} obtained 
benchmark SDCS using an external complex scaling technique
\cite{MRB97} that
does not require the knowledge of three-body boundary
conditions. These were found to be in consistent agreement with the
CCC results. Furthermore,
\citeasnoun{RMIB99} showed
how step functions may arise when discretization with short-ranged
potentials is used. 

To our mind the closest to a proof of the step function idea has 
been given by \citeasnoun{S99l}. He showed that the close-coupling
equations, obtained by using exact target eigenstates to expand the
total wave function, have unitarity satisfied with the secondary energy 
integration ending at $E/2$. This implies a step function in the
underlying amplitudes since the coupled equations are formally written 
with this integration ending at $E$. Given that the CCC
square-integrable target states
form an equivalent quadrature rule for the infinite summation over the true
target discrete eigenstates simultaneously with an integration over
the true target continuum it is tempting to conclude that for infinite 
$N$ the CCC equations converge to those obtained using exact target
eigenstates, and hence the CCC ionization amplitudes should display a step
function behaviour. Furthermore, 
by comparison with the known SDCS at $E/2$, 
he observed that the CCC-calculated singlet SDCS appeared to converge to $1/4$ 
the value of the true result, and suggested that the CCC equations
appeared to behave like Fourier expansions of the underlying
amplitudes. A Fourier expansion of a step function converges to the 
midpoint of the step height. Therefore, the CCC amplitude at
equal energy sharing converges to $1/2$ of the step height, and hence 
the SDCS to $1/4$ of the true height.

This interpretation is very exciting because it explains the apparent
convergence of the SDCS at $E/2$, even when convergence is lacking at
unequal energy-sharing, and how it may be related to the true 
result. A detailed set of applications to the calculation
of equal-energy-sharing fully differential electron-impact ionization
of the atomic hydrogen ground state has been given \cite{B99jpb}. Here we
examine convergence for scattering from the excited states, and
particularly of the free-free transitions. The latter are
interesting because it is the free-free $V$-matrix
elements that are responsible for the failure to date of solving the
close-coupling 
equations involving pure atomic (discrete and continuous) eigenstates,
and thereby requiring the 
introduction of a pseudostate approach. Free-free one-electron
transitions have been 
looked at before, see \citeasnoun{CF99} for example. Here, for the
first time to the best of our knowledge, free-free transitions
involving two electrons are shown to be calculable.

Since we shall only concern ourselves with the S-wave model, momenta
will be written as scalars in what follows.
The traditional close coupling equations arise upon expanding the total wave
function  over the complete set of target eigenstates $\phi_n$ of
energy $\epsilon_n$. Though we use a discrete notation,
this involves an infinite sum of the bound states
$\phi_n(\epsilon_n)$ and an integral ($d\epsilon$)
over the continuum states $\phi(\epsilon)$. The close-coupling
equations may be written  
as coupled Lippmann-Schwinger equations for the $T$-matrix \cite{BS92}
\begin{eqnarray}
\langle {k}_f\phi_f|T_S|\phi_i{k}_i\rangle &=&
\langle {k}_f\phi_f|V_S|\phi_i{k}_i\rangle \nonumber\\
&&+\isum_nd\epsilon_n\int dk\frac{\langle {k}_f\phi_f|V_S|\phi_n{k}\rangle
\langle {k}\phi_n|T_S|\phi_i{k}_i\rangle}
{E+i0-\epsilon_n-k^2/2}.
\label{TLS}
\end{eqnarray}
These equations are yet to be solved directly due to the non-existence of 
the free-free matrix elements $\langle
k'\phi(\epsilon'')|V_S|\phi(\epsilon)k\rangle$. We write the cross sections for
the discrete transition $i\to f$ as
\begin{equation}
\sigma_{fi}^{(S)}=\frac{k_f}{k_i}
|\langle {k}_f\phi_f|T_S|\phi_i{k}_i\rangle|^2,
\label{disc}
\end{equation}
and for an ionization process as
\begin{equation}
\sigma_{i}^{(S)}(\epsilon)=\frac{k}{\sqrt{2\epsilon}k_i}
|\langle {k}\phi(\epsilon)|T_S|\phi_i{k}_i\rangle|^2.
\label{cont}
\end{equation}
Then the total cross section $\sigma^{(S)}$, at energies above the
ionization threshold ($E>0$), for scattering from some
initial state $i$ is
\begin{equation}
\sigma^{(S)}_i= \sum_{f=1}^\infty \sigma_{fi}^{(S)}
+\int_0^E \sigma_{i}^{(S)}(\epsilon)d\epsilon.
\label{tot}
\end{equation}
The continuum integration ending at E comes from the fact that  in
\eref{TLS} on the
energy shell $\epsilon_n\le E$. From \eref{tot} we see
immediately the fundamental problem of the close-coupling
equations. Since antisymmetry is explicitly included in the $V_S$
\cite{BS92} there appears to be a double-counting problem as the
energy integration ends at $E$ and not $E/2$. However, as mentioned above,
\citeasnoun{S99l} has shown that there is no contribution to the
total cross section from $\langle {k}\phi(\epsilon)|T_S|\phi_i{k}_i\rangle$
for $\epsilon>k^2/2$ thereby reducing the integration endpoint to $E/2$ and
bringing about consistency with formal ionization theory \cite{R68}.

In order to solve \eref{TLS} the CCC method uses $N$ discrete states
$\phi_n^{(N)}$, with energies $\epsilon_n^{(N)}$, obtained by
diagonalising the target Hamiltonian in an orthogonal
Laguerre basis \cite{BS92}. The coupled Lippmann-Schwinger equations
then take the form
\begin{eqnarray}
\fl\langle {k}_f\phi_f^{(N)}|T_S^{(N)}|\phi_i^{(N)}{k}_i\rangle &=&
\langle {k}_f\phi_f^{(N)}|V_S^{(N)}|\phi_i^{(N)}{k}_i\rangle \nonumber\\
&&+\sum_{n=1}^N\int dk\frac{\langle {k}_f\phi_f^{(N)}|V_S^{(N)}|
\phi_n^{(N)}{k}\rangle
\langle {k}\phi_n^{(N)}|T_S^{(N)}|\phi_i^{(N)}{k}_i\rangle}
{E+i0-\epsilon_n^{(N)}-k^2/2}.
\label{TLSN}
\end{eqnarray}
Using the relation 
\begin{equation}
|\phi(\epsilon_f)\rangle=\lim_{N\to\infty}|\phi_f^{(N)}\rangle
\langle\phi_f^{(N)}|\phi(\epsilon_f)\rangle,
\end{equation}
where $\epsilon_f^{(N)} = \epsilon_f$, the total cross section
corresponding to  \eref{tot} becomes
\begin{equation}
\sigma^{(SN)}_i= \sum_{f:\epsilon_f^{(N)}<0} \sigma_{fi}^{(SN)}
+\int_0^E \sigma_{i}^{(SN)}(\epsilon)d\epsilon,
\label{totN}
\end{equation}
where
\begin{equation}
\sigma_{i}^{(SN)}(\epsilon_f) = \frac{k_f}{\sqrt{2\epsilon_f}k_i}
|\langle\phi(\epsilon_f)|\phi_f^{(N)}\rangle
\langle {k}_f\phi_f^{(N)}|T_S^{(N)}|\phi_i^{(N)}{k}_i\rangle|^2
\label{sdcs}
\end{equation}
is the SDCS. For infinite $N$ \eref{totN} goes to \eref{tot} and hence 
a step function, with the integration ending effectively at $E/2$.

Similarly, we can write down the relationship 
between the free-free matrix elements occurring in both \eref{TLS} and
\eref{TLSN}. For example,
\begin{equation}
\fl\langle k_f\phi(\epsilon_f)|V_S|\phi(\epsilon_i)k_i\rangle
=\lim_{N\to\infty}\langle\phi(\epsilon_f)|\phi_f^{(N)}\rangle
\langle {k}_f\phi_f^{(N)}|V_S^{(N)}|\phi_i^{(N)}{k}_i\rangle 
\langle\phi(\epsilon_i)|\phi_i^{(N)}\rangle.
\end{equation}
Thus, the
non-existence of free-free $V_S$ matrix elements in \eref{TLS} has not been
eliminated, and becomes evident with increasing $N$. However,
numerical solutions of \eref{TLSN} have shown good convergence for the 
$T_S$ matrix elements, at least for excitation of the ground state
\cite{BS92l}. Here we check for convergence in the case of excited
initial states including a free-free transition.

\begin{figure}
\hspace{-1truecm}\epsffile{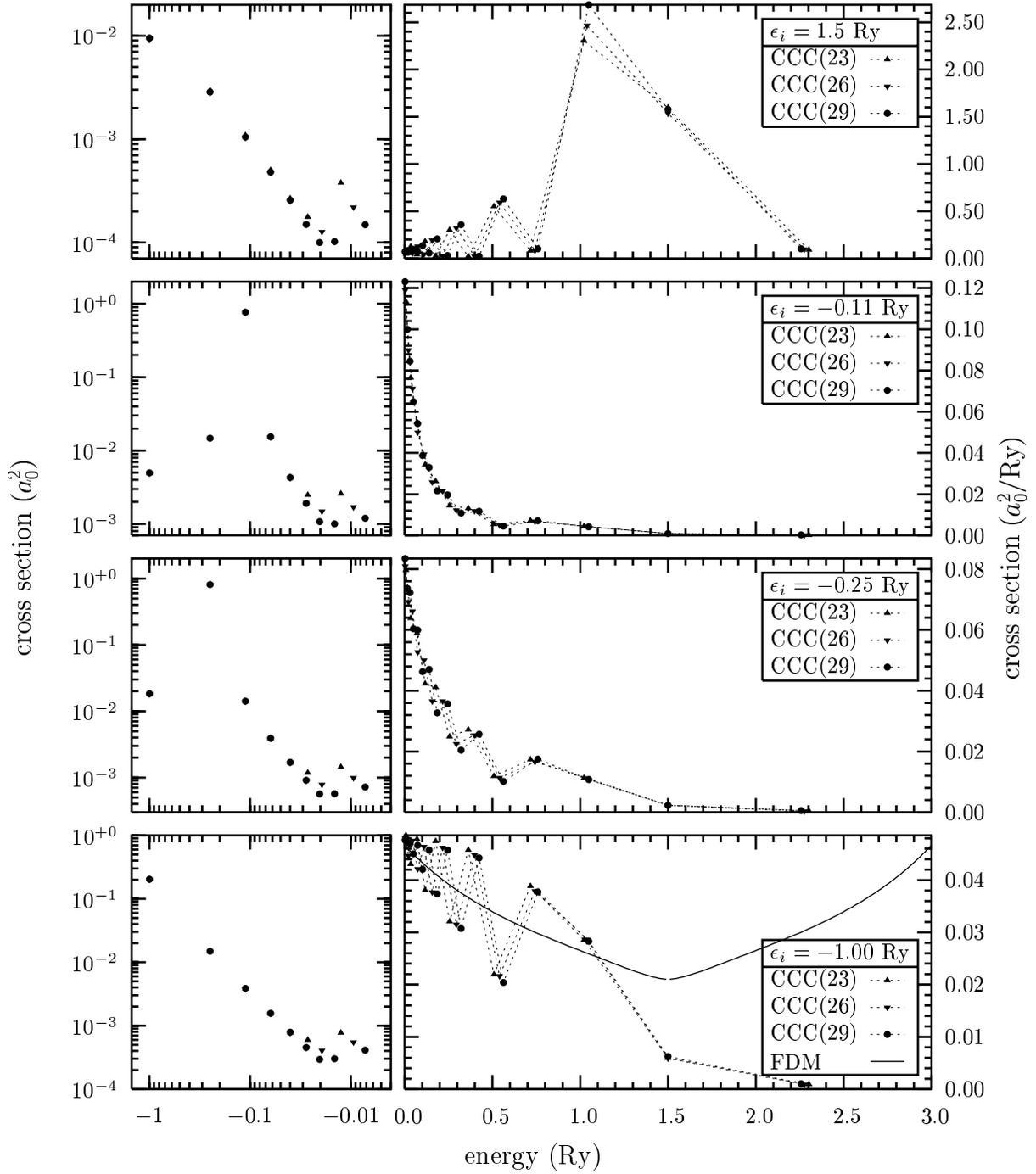}
\caption{The singlet cross sections arising upon solution of the
electron-hydrogen S-wave model at the total energy of 3~Ry for the
lowest three discrete (1S, 2S and 3S) initial state, and the
$\epsilon_i=1.5$~Ry state from the target continuum. The present
CCC($N$) calculations are described in the text. The SDCS calculated
by the  finite-difference 
method of \protect\citeasnoun{JS99l} is denoted by FDM.
}
\label{fig1}
\end{figure}

The numerical investigation is performed for the total energy
$E=3$~Ry. The results of three calculations, $N$=23, 26 and 29, are
presented. The states were chosen in such a way so that there was
always a state of 1.5~Ry. This way all three calculations contain the
matrix elements of the free-free transition
corresponding to two 1.5~Ry electrons elastically scattering on a proton.
In \fref{fig1}
we present the discrete excitation cross sections and the SDCS,
evaluated according to \eref{sdcs}, for the
singlet case. 
The value $\epsilon_i$ is the initial energy of the bound electron
when negative, or otherwise the energy of an
incident electron.

We begin the discussion of the cross sections for the negative-energy
states. Good convergence is seen for the first five states for all
four initial states, with elastic scattering being the most dominant.
For the higher ($n>5$) lying discrete states the bigger 
calculations yield the smaller cross sections, but in all cases the
cross sections for the last negative-energy states rise. This is not
an indication of divergence from the expected $n^{-3}$ scaling rule,
but shows how the least negative-energy states take into account the remaining
full infinite discrete spectrum.

Turning our attention to the SDCS from the ground state, for
energies less than $E/2$ we observe that there are
substantial $N$-dependent oscillations about the exact result,
calculated using the finite difference method (FDM) by
\citeasnoun{JS99l}. At $E/2$ the three CCC
calculations show convergence to approximately a quarter of the FDM
result, as expected.

The SDCS from the 2S and 3S initial states show less oscillation than
for the ground state owing to the SDCS at $E/2$ being of relatively
small magnitude. Thus, within the same calculations the CCC method is
able to obtain SDCS more accurately, over the energy range $[0,E/2]$,
from excited states than from the ground state.

Finally, we consider the free-free transitions for the case where the
two electrons are both incident at 1.5~Ry. It is seen that the
functional form of the SDCS changes as compared to the discrete
initial states.  Oscillations are very large, but convergence at $E/2$ 
is evident, and presumably to one quarter of the true value. This
suggests that the elastic scattering is the most dominant, which explains
the functional form change, and is consistent with the elastic scattering
from the presented discrete states being the most dominant of the
discrete transitions. It is truly remarkable to see convergence at
$E/2$ as the corresponding $V_S$ matrix elements are an order of
magnitude greater than the $T_S$ matrix elements and continue to
increase with $N$.

\begin{figure}
\hspace{-1truecm}\epsffile{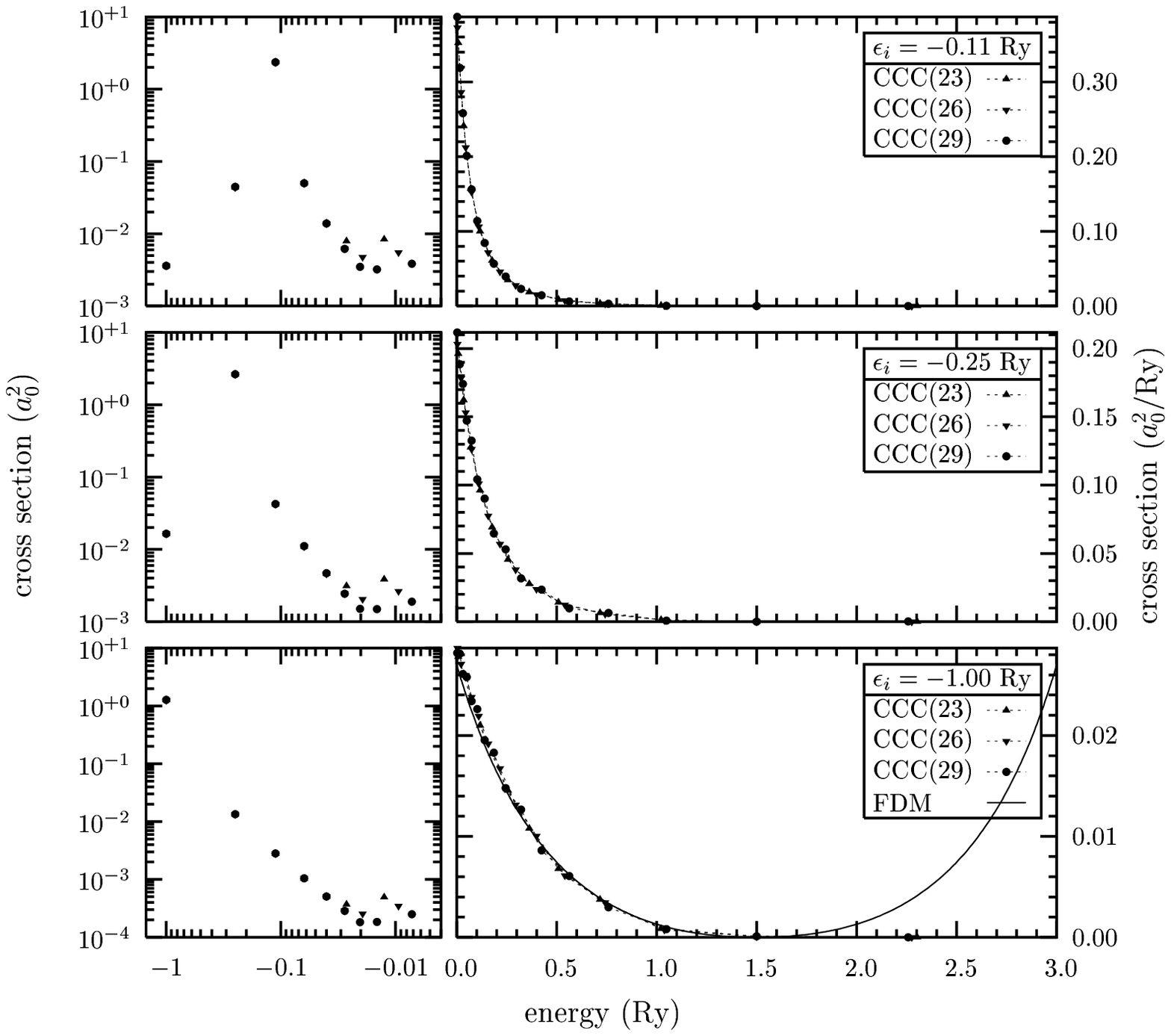}
\caption{Same as for \protect\fref{fig1} except for the triplet
case. No result for scattering from the $\epsilon_i=1.5$~Ry state is
given owing to the Pauli Principle ensuring that such cross sections
are zero.}
\label{fig2}
\end{figure}

For completeness,
in \fref{fig2} we present the cross sections for the triplet case. Here the
initial state with two 1.5~Ry electrons is forbidden and so is not
presented. All convergence considerations for the discrete excitations 
apply equally here as in the case of singlet scattering. The SDCS are
all free from oscillations owing to the zero cross section at $E/2$,
and good agreement is found with the FDM-calculated SDCS available
only for the ground state \cite{JS99l}.

In summary, the recent work of \citeasnoun{S99l} has shown that the
CCC theory yields convergent ionization scattering amplitudes at equal
energy-sharing that are simply a factor of two less than the true
amplitudes. Thus, the CCC theory may claim to yield these amplitudes
accurately for any initial state, and not only for the S-wave model
considered here. We have seen convergence in the model for the elastic 
free-free transition which corresponds to the real experimental case
of equal energy (2e,2e) on a proton. Whereas such processes are yet to 
be experimentally observed the CCC (e,2e) calculations include such
processes as an intermediate step and these may be extracted as
convergent cross sections. This is particularly pleasing since the
introduction of the $L^2$ technique in solving the close-coupling
equations does not eliminate the divergence of the underlying
free-free potential matrix elements, but masks it with a dependence on 
$N$. Finally, though \citeasnoun{S99l} does not
claim this, we suggest that his work implies a step function of the
underlying amplitudes in forming \eref{tot} and hence the
CCC-calculated amplitudes used in \eref{totN}, supporting our initial
hypothesis \cite{B97l}.

The authors thank Andris Stelbovics for many discussions and
communication of results prior to publication. The support of the
Australian Research Council and the Flinders University is acknowledged.

\section*{References}

\end{document}